\documentclass[review]{elsarticle}

\usepackage{amsmath}
\usepackage{amssymb}

\usepackage{soul,color}
\usepackage{textcomp}

\usepackage{lineno,hyperref}
\modulolinenumbers[1]

\newcommand{\tr}[1]{\mathrm{Tr} \left\{ #1 \right\}}
\newcommand{\sx}{I_\mathrm{x}}
\newcommand{\sy}{I_\mathrm{y}}
\newcommand{\sz}{I_\mathrm{z}}
\newcommand{\hdz}{H_\mathrm{dz}}

\begin{document}

\begin{frontmatter}

\title{Many-spin entanglement in multiple quantum NMR with a dipolar ordered initial state}

\author[icp]{E.~B.~Feldman}
\author[icp,msu]{I.~D.~Lazarev} %

\address[icp]{Institute of Problems of Chemical Physics of Russian Academy of Sciences, \\ Chernogolovka, Moscow Region, Russia 142432}
\address[msu]{Faculty of Fundamental Physical-Chemical Engineering, Lomonosov Moscow State University, GSP-1, Moscow, Russia 119991}

\begin{abstract}
Multiple quantum (MQ) NMR with a dipolar ordered initial state opens new possibilities for the exploration of many-spin entanglement. 
In this paper, we investigate many-spin entanglement in a gas of spin-carrying molecules (atoms) in nanocavities 
in the conditions of MQ NMR with a dipolar ordered initial state.
The second moment of the distribution of the intensities of MQ NMR coherences,
which provides a lower bound on the quantum Fisher information, 
is used for an estimate of the number of the entangled spins. 
Many-spin entanglement is investigated at different temperatures and different numbers of spins.
\end{abstract}


\end{frontmatter}


\section{Introduction}
\label{sec:1}

Entanglement~\cite{Nielsen_2009} is an important notion in quantum mechanics. It is responsible, in particular, for advantages of quantum computers over their classical counterparts.
Recently demonstrated quantum supremacy of a programmable superconducting processor~\cite{Arute2019} is also connected with entanglement, which is absent in classical physics.
Among numerous methods of the investigation of entanglement, we focus on multiple quantum (MQ) NMR in solids~\cite{Baum_1985}, which is widely used to characterize entanglement in binary systems~\cite{Furman_2008,Furman_2009,Fel_dman_2008,Fel_dman_2012}. 
It turns out that the MQ NMR spectroscopy~\cite{Baum_1985} allows us to extract information about many-spin entanglement~\cite{G_rttner_2018} by using the quantum Fisher information~\cite{T_th_2014,Pezz__2018}.

The quantum Fisher information describes the quickness of a change of quantum states determined by a density matrix in response to a change of some parameter. 
In MQ NMR spectroscopy, that parameter is the phase increment between the radio-frequency (rf) pulses, irradiating the system on the preparation and mixing periods of the MQ NMR experiment~\cite{Baum_1985} 
that leads to the separation of the signals, corresponding to MQ NMR coherences of different orders, and allows obtaining the total MQ NMR spectrum. 
The phase increment is proportional to the duration of the evolution period.
Using the quantum Fisher information (QFI)~\cite{Liu_2014} for the analysis of the MQ NMR spectra, one can extract important information about many-spin entanglement.
The point is that there is a relation between the second moment of the MQ NMR spectrum~\cite{Khitrin_1997} and the quantum Fisher information~\cite{G_rttner_2018,Doronin_2019}.
Moreover, the second moment of the MQ NMR spectrum provides a lower bound on the quantum Fisher information~\cite{G_rttner_2018}.
This means that the MQ NMR spectroscopy is a valuable method for solving quantum information problems.

Many-spin entanglement was investigated~\cite{Doronin_2019} for a nonspherical nanopore filled with a gas of spin-carrying molecules in a strong external magnetic field~\cite{Baugh_2001,Doronin_2009}.
The thermally equilibrium initial state of the system was determined by the one-spin Zeeman interaction with the external magnetic field~\cite{Doronin_2007a}.
It is also possible to investigate many-spin entanglement when the same system is initially prepared in the dipolar ordered state~\cite{Goldman_1970} using either the adiabatic demagnetization method in a rotating reference frame (RRF)~\cite{Goldman_1970,Slichter_1961} or the two-pulse Broekaert-Jeener sequence~\cite{Goldman_1970,Jeener_1967}.
The MQ NMR dynamics with this initial state have been simulated both in small spin systems~\cite{Doronin_2007a,Doronin_2007b} and in a system consisting of 200-600 spin-carrying molecules (atoms) filling a nanopore~\cite{Doronin_2011}.
The approaches developed for those investigations are restricted to  high temperatures and cannot be applied to  many-spin entanglement.

In the present article we consider the intermediate-temperature case with low Zeeman temperatures and high dipolar ones. 
Nuclear magnetic ordering~\cite{Abragam_1982} is beyond the scope of this paper.
Notice that the two-pulse Broekaert-Jeneer experiment~\cite{Jeener_1967} was performed in the high-temperature case.
We prove theoretically that the experiment~\cite{Jeener_1967} can be realized also for the intermediate-temperature case. 
It was shown~\cite{Doronin_2011} that in the MQ NMR experiment with the dipolar ordered initial state, MQ NMR coherences emerge faster 
than in the MQ NMR experiment with the thermal-equilibrium initial state in a strong external magnetic field.
This observation is important for many-spin entanglement investigations, because they involve calculations of the second moment of the distribution of MQ NMR coherences. 
It is also important for the investigation of correlation spreading~\cite{Baugh_2001,Baum_1986,S_nchez_2014,Munowitz_1987} and localization~\cite{Alvarez_2015,Wei_2018}.
Indeed, the spreading rate can be described through out-of-time ordered correlations, which are connected with the distribution of MQ NMR coherences. 

The present paper investigates many-spin entanglement using the MQ NMR spectrum of spin-carrying atoms (molecules) in a nanopore when the system is prepared in a dipolar ordered state.
In Sec.~\ref{sec:2}, the theory of MQ NMR dynamics at a low Zeeman temperature and a high dipolar temperature is developed.
An analytical solution for the MQ NMR dynamics of a three-spin system is obtained at such temperatures in Sec.~\ref{sec:3}.
The second moment of the MQ NMR spectrum as a measure of many-spin entanglement is considered in Sec.~\ref{sec:4}.
The dependence of many-spin entanglement on the dipolar temperature and the number of the spins in the system is investigated in Sec.~\ref{sec:5}.
We briefly summarize our results in concluding Sec.~\ref{sec:6}.
In the Appendix, we show that the two-pulse Broekaert-Jeener sequence can be used in the case when the Zeeman temperature is low and the dipole one is high.

\section{Theory of MQ NMR dynamics in a nanopore at a low Zeeman temperature and a high dipolar temperature}
\label{sec:2}

MQ NMR dynamics in a nanopore is governed by the Hamiltonian~\cite{Doronin_2019,Doronin_2009} 
\begin{equation}
    \label{eq:1}
    H_{\mathrm{MQ}} = - \dfrac{D}{4} \left[
        \left(I^{+}\right)^{2} 
        + \left(I^{-}\right)^{2} 
    \right] ,
\end{equation}
where 
\begin{equation}
    \label{eq:2}
    I^{\pm} = \sum\limits_{j=1}^{N} I_{j}^{\pm},
\end{equation}
$N$ is the number of the spins in the nanopore, $I^{\pm}_{j}$ are the raising or lowering operators of spin $j$, and $D$ is the dipolar coupling constant averaged by the fast molecular diffusion of spin-carrying atoms (molecules) in the nanopore.
We emphasize that the dipolar coupling constant $D$ is the same for all pairs of interacting spins in the nanopore~\cite{Doronin_2019,Doronin_2009}.
The density matrix $\rho(\tau)$ on the preparation period of the MQ NMR experiment~\cite{Baum_1985} can be obtained from the Liouville evolution equation~\cite{Goldman_1970,Abragam_1982} 
\begin{equation}
    \label{eq:3}
    i\dfrac{\mathrm{d}\rho(\tau)}{\mathrm{d}\tau} = \left[
    H_\mathrm{MQ},\rho(\tau)
    \right]
\end{equation}
with the initial thermodynamic equilibrium density matrix 
\begin{equation}
    \label{eq:4}
       \rho(0) = \rho_\mathrm{eq} = \dfrac{1}{Z}
       e^{
		   \frac{\hslash \omega_{0}}{k} \alpha_\mathrm{z} I_\mathrm{z} 
            + \frac{\hslash }{k} \beta_\mathrm{d} H_\mathrm{dz}
        },
\end{equation}
where 
$Z = \mathrm{Tr} \left\{ e^{\frac{\hslash \omega_{0}}{k} \alpha_\mathrm{z} I_\mathrm{z} + \frac{\hslash  }{k} \beta_\mathrm{d} H_\mathrm{dz}} \right\}$ is the partition function, 
$\hslash$ and $k$ are the Plank and Boltzmann constants, 
$\omega_{0}$  is the Larmor frequency , $I_\mathrm{z}$ is the operator of the projection of the total spin angular momentum on the z-axis, 
which is directed along the strong external magnetic field,  
$H_\mathrm{dz}$ is the secular part of the dipole-dipole interaction~(DDI) Hamiltonian in a strong external magnetic field, and $\alpha_\mathrm{z}$, $\beta_\mathrm{d}$ are the inverse Zeeman and dipolar temperatures. 
We will consider the case when the Zeeman temperature is low $({\frac{\hslash \omega_{0}}{k} \alpha_\mathrm{z}}\gg 1)$ 
and the dipolar temperature is high $\left( \frac{\hslash{D}}{k}\beta_\mathrm{d} \ll 1\right)$.
For concreteness, we suppose that $\omega_{0} = 2\pi \cdot 500 \cdot 10^{6}$~s$^{-1}$ and $D = 2\pi \cdot 10^{4}$~s$^{-1}$.
In the Appendix, we prove that the two-pulse Broekaert-Jeener sequence~\cite{Goldman_1970,Jeener_1967} results in the dipolar ordered state even at a low Zeeman temperature.
The adiabatic demagnetization~\cite{Goldman_1970,Slichter_1961} is the second method of preparing a system in the dipolar ordered state.
Using those methods, we can obtain the system in the thermodynamic equilibrium state with the density matrix
\begin{equation}
    \label{eq:5}
    \rho_i = \frac{1}{Z_i} e^\frac{\hslash\beta_\mathrm{d} \hdz}{k}
    \approx \frac{1}{Z_i}(1 + \frac{\hslash\beta_\mathrm{d}}{k} H_\mathrm{dz}),
\end{equation}
where the partition function
\begin{equation}
    \label{eq:6}
	Z_i = \mathrm{Tr} \left\{ e^\frac{\hslash\beta_\mathrm{d} \hdz}{k} \right\} \approx 2^{N}.
\end{equation}
MQ NMR dynamics in the nanopore will be investigated on the basis of Eq.~(\ref{eq:3}) with the initial state of Eq.~(\ref{eq:5}).
It is also significant that the Hamiltonian $H_{dz}$ is partially averaged by the fast molecular diffusion in the nanopore and the averaged Hamiltonian can be written as \cite{Fel_dman_2004,Doronin_2011}
\begin{equation}
    \label{eq:7}
    H_\mathrm{dz} = \dfrac{D}{2} (3 I^{2}_{z} - I^{2}) , 
\end{equation}
where $I^{2}$ is the square of the spin angular momentum.

Let   $G(\tau,\phi)$  be the signal after the preparation, evolution and mixing periods of the MQ NMR experiment~\cite{Baum_1985}, averaged over the equilibrium density matrix.  $G(\tau,\phi)$ can be written as~\cite{Doronin_2019} 
\begin{equation}
    \begin{split}
        \label{eq:8}
        G(\tau,\phi) 
        & = \mathrm{Tr}\left\{
            e^{i H_\mathrm{MQ} \tau} e^{i\phi I_\mathrm{z}} e^{-i H_\mathrm{MQ}\tau} 
            \rho_i 
            e^{i H_\mathrm{MQ} \tau} e^{-i \phi I_\mathrm{z}} e^{-i \phi H_\mathrm{MQ} \tau} 
            \rho_i 
        \right\} \\
        & = \mathrm{Tr} \left\{
        e^{i \phi I_\mathrm{z}}
        \rho(\tau) 
        e^{-i \phi I_\mathrm{z}} 
        \rho(\tau) 
        \right\},
    \end{split}
\end{equation}
where
\begin{equation}
    \label{eq:9}
    \rho(\tau) 
    = e^{-i H_\mathrm{MQ} \tau } 
    \rho_i 
    e^{i H_\mathrm{MQ} \tau}
\end{equation}
is the solution of Eq.~(\ref{eq:3}) at the initial condition of Eq.~(\ref{eq:5}).
It is convenient to expand the spin density matrix, $\rho(\tau)$, in series as
\begin{equation}
    \label{eq:10}
    \rho(\tau) = \sum\limits_n \rho_n(\tau),
\end{equation}
where $\rho_{n}(\tau)$ is the contribution to $\rho(\tau)$ from the MQ coherence of the $n$-th order~\cite{Fel_dman_1996}.
Then the function $G(\tau,\phi)$ of Eq.~(\ref{eq:8}) can be rewritten as 
\begin{equation}
    \label{eq:11}
    G(\tau,\phi) 
    = \sum\limits_n e^{i n \phi} \mathrm{Tr} \left\{ 
        \rho_{n}(\tau) \rho_{-n}(\tau) 
    \right\},
\end{equation}
where we took into account that
\begin{equation}
    \label{eq:12}
    \left[ I_{\mathrm{z}},\rho_n(\tau) \right] = n \rho_n(\tau)
\end{equation}
It is necessary for further calculations to introduce the normalized intensities $J_{n}(\tau)$ $(n=0, \pm 2, \pm 4, \cdots)$ of the MQ NMR coherences
\begin{equation}
    \label{eq:13}
    J_{n}(\tau) = \dfrac{\mathrm{Tr} \left\{
    \rho_{n}(\tau) \rho_{-n}(\tau) 
    \right\}} 
    {\mathrm{Tr} \left\{\rho^2_{i} \right\}}
\end{equation}
Using Eqs.~(\ref{eq:9}),~(\ref{eq:10}) one can verity that 
\begin{multline}
    \label{eq:14}
    \sum\limits_{n} J_{n}(\tau)
    = \dfrac{
        \mathrm{Tr} \left\{
            \sum_{n} \rho_{n}(\tau) \rho_{-n}(\tau)
        \right\}}
    {\mathrm{Tr} \left\{ \rho^2_{i} \right\}} 
    = \dfrac{
        \mathrm{Tr} \left\{
            \sum_{\mathrm{m,n}} \rho_n(\tau)\rho_m(\tau)
    \right\}}
    {\mathrm{Tr} \left\{\rho^2_{i}\right\}}
    \\
    = \dfrac{
        \mathrm{Tr}\left\{\rho^2(\tau)\right\}
    }
    {
        \mathrm{Tr}\left\{\rho^2_{i}(\tau)\right\}
    }
    = \dfrac{
        \mathrm{Tr} \left\{ 
            e^{-i H_\mathrm{MQ} \tau} 
            \rho^{2}_{i}
            e^{i H_\mathrm{MQ} \tau} 
        \right\}
    }
    {
        \mathrm{Tr} \left\{ \rho_{i}^{2} \right\}
    } 
    = 1
\end{multline}
One can conclude from Eq.~(\ref{eq:14}) that the sum of the MQ NMR coherences is conserved on the preparation period of the MQ NMR experiment~\cite{Baum_1985}.

The basis consisting of the eigenstates of the operator $I_\mathrm{z}$ (dubbed the multiplicative basis) is widely used for numerical calculations of MQ NMR dynamics~\cite{Zhang_2009}.
Due to the rapid expansion of the Hilbert space with the growth of the number of spins such calculations are possible only for systems with a small number of spins.
That approach is not suitable for investigations of many-spin entanglement.
Since the Hamiltonian $H_{MQ}$ of Eq.~(\ref{eq:1}) commutes with the square of the total spin angular momentum $\hat I^2$, 
it is possible to use the basis consisting of the common eigenstates of $\hat I^2$ and $I_\mathrm{z}$ in order to study MQ NMR dynamics as was done in Ref.~\cite{Doronin_2009,Doronin_2011,Doronin_2019}.
In this basis, the Hamiltonian $H_{MQ}$ and the initial density matrix of Eq.~(\ref{eq:5}) (see also Eq.~(\ref{eq:7})) consist of blocks, corresponding to different values of the spin angular momentum~\cite{Doronin_2009}.
Then the investigation of MQ NMR dynamics can be reduced to solving a set of problems of lower dimensions.

Since the Hamiltonian $H_{MQ}$ of Eq.~(\ref{eq:1}) commutes with the operator $e^{i\pi I_\mathrm{z}}$, the $2^N \times 2^N$ Hamiltonian matrix reduces to two $2^{N-1} \times 2^{N-1}$ submatrices~\cite{Doronin_2009}.
For odd $N$, both submatrices give the same contribution to the MQ NMR coherences, and one should solve the problem using only one $2^{N-1} \times 2^{N-1}$ submatrix and double the obtained intensities. 
In our calculations, we take only odd numbers of spins.
Using this method, one can investigate MQ NMR dynamics in systems consisting of hundreds spins.

\section{Analytical solution for MQ NMR dynamics of a three-spin system in a nanopore in a dipolar ordered state}
\label{sec:3}

\begin{figure}
    \centering
  	\includegraphics[width=0.5\linewidth]{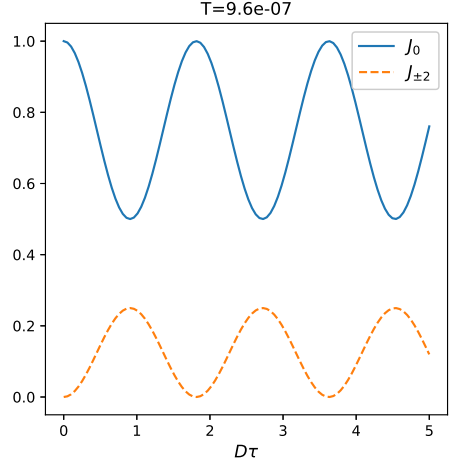}
	\caption{
	    Intensities of MQ NMR coherenes $J_{n}$ ($n=0, 2$) in a nanopore with $N=3$.
	}
	\label{fig:1}
\end{figure}

Obtaining the exact solution for MQ NMR dynamics of a three-spin system in a dipolar ordered state in a nanopore is similar to the problem considered in Ref.~\cite{Doronin_2019} for the initial thermodynamic equilibrium in a strong external magnetic field. 
Here we do not use the high temperature approximation~\cite{Goldman_1970}.

The Hamiltonian $H_{MQ}$ of Eq.~(\ref{eq:1}) consists here of two blocks for the two possible values of the spin angular momentum $(I^2 = S(S+1),  \quad S=3/2,1/2)$.
Those blocks and the corresponding eigenvalues and eigenstates are given in Ref.~\cite{Doronin_2019}.
The density matrix of the system consists also of two blocks $\rho^{3/2}(\tau)$, $\rho^{1/2}(\tau)$, and 
\begin{equation}
    \label{eq:15} 
    \rho^{3/2}(0) = \dfrac 1 Z
    \begin{pmatrix}
        e^{\frac{3b}{2}} & 0 & 0 & 0 
        \\
        0 & e^{\frac{-3b}{2}} & 0 & 0 
        \\
        0 & 0 & e^{-\frac{-3b}{2}} & 0 
        \\
        0 & 0 & 0 & e^{\frac{3b}{2}}
    \end{pmatrix}, 
    \quad
    \rho^{1/2}(0) = \dfrac 1 Z
    \begin{pmatrix}
       	1 & 0 
        \\
        0 & 1
    \end{pmatrix}
\end{equation}
where $b = \dfrac{\hslash D}{k\mathrm{T}}$and $T$ is the temperature.
After simple calculations one can obtain the density matrices $\rho^{3/2}(\tau)$ and $\rho^{1/2}(\tau)$, 
which allow us to find the intensities of the MQ NMR coherences.

Only the MQ NMR coherences of the zeroth and plus/minus second orders appear in the considered systems. 
The intensities of these coherences are
\begin{equation}
    \begin{split}
        \label{eq:16}
        J_0(\tau) & = 1 
        - \dfrac 1 2 \tanh^2\left( \dfrac{3b}{2} \right)
            \sin^2 \left( \sqrt{3} Dt \right), 
        \\
        J_{\pm2}(\tau) & = \dfrac{1}{4} 
            \tanh^2 \left( \dfrac{3b}{2} \right)
            \sin^2 \left( \sqrt{3} Dt \right)
    \end{split}
\end{equation}
The sum of the intensities of Eq.~(\ref{eq:16}) equals one in accordance with Eq.~(\ref{eq:14}).
The dependencies of the calculated intensities $J_{n}(\tau)$ $(n=0,2)$ on the evolution time are shown in Fig.~(\ref{fig:1}).

\section{Second moment of the MQ NMR spectrum as a measure of many-spin entanglement}
\label{sec:4}

The expression~(\ref{eq:8}) for the MQ NMR signal $G(\tau,\phi)$ can be expanded in series in the phase increment $\phi$:
\begin{equation}
    \begin{split}
        \label{eq:17}
        G(\tau,\phi)  
        & = \mathrm{Tr} \left\{ 
            \rho(\tau) e^{i \phi I_\mathrm{z} }
            \rho(\tau) e^{-i\phi I_\mathrm{z}}
        \right\}  \\
        & = \mathrm{Tr} \left\{ \rho^2(\tau) \right\} 
        - \phi^2 \mathrm{Tr} \left\{ 
            \rho^2(\tau) I^2_\mathrm{z} 
            - (\rho(\tau) I_\mathrm{z})^2
        \right\} 
        + O(\phi^3)
    \end{split}
\end{equation}
It is possible to prove~\cite{Girolami_2017} that the quantum Fisher information $F_\mathrm{Q}(\rho,I_\mathrm{z})$~\cite{Helstrom_1976}
\begin{equation}
    \label{eq:18}
    F_\mathrm{Q}(\rho,I_\mathrm{z}) \geq 4 \mathrm{Tr} \left\{ \rho^2 I^2_\mathrm{z} - (\rho I_\mathrm{z})^2 \right\}
\end{equation}
At the same time, it is easy to verify that $2 \mathrm{Tr} \left\{ \rho^2(\tau) I_\mathrm{z}^2 - \left( \rho(\tau) I_\mathrm{z} \right)^2 \right\}$ equals the second moment $M_2$ of the distribution of the intensities of the MQ NMR coherences~\cite{Khitrin_1997}
\begin{equation}
    \label{eq:19}
    M_2 = \sum_{n} n^2 J_n (\tau) ,
\end{equation}
where $J_n(\tau)$ ($n=0,\pm 2, \pm 4, \cdots$) is determined by Eq.~(\ref{eq:13}).
Thus, the second moment of the MQ NMR spectrum provides a lower bound on the quantum Fisher information $F_\mathrm{Q}(\rho,I_\mathrm{z})$.
It was also shown~\cite{T_th_2014,Pezz__2018} that if
\begin{equation}
    \label{eq:20}
    F_\mathrm{Q} (\rho,I_\mathrm{z}) > n k^2 + (N - n k)^2,
\end{equation}
where $n$ is the integer part of ${N/k}$, then the system with the density matrix $\rho(\tau)$ is $(k+1)$-spin entangled~\cite{Pezz__2009,Hyllus_2012,T_th_2012}.
The results of the numerical analysis of many-spin entanglement in the system of spin-carrying molecules (atoms) initially prepared in the dipolar ordered state are presented in the following section.

\section{Numerical analysis of many-spin entanglement at different temperatures and numbers of spins in the system}
\label{sec:5}

\begin{figure}
  	\includegraphics[width=0.95\linewidth]{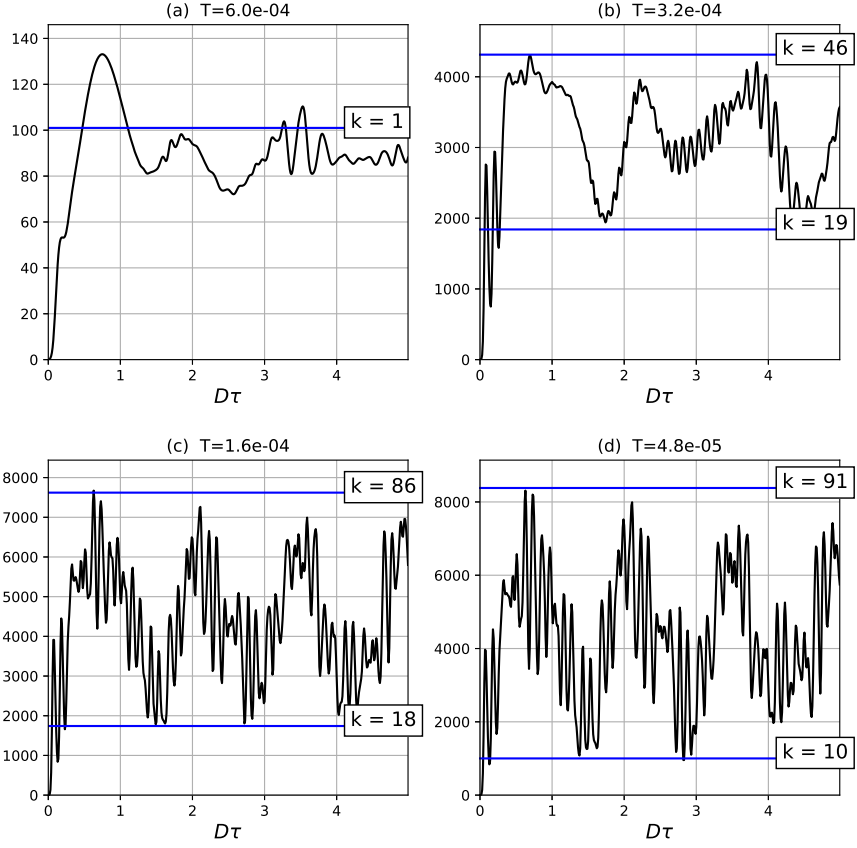}
	\caption{
	    The dependence of the lower bound on the quantum Fisher Information $F_\mathrm{Q} = 2 M_{2}$ 
	    on the dimensionless time $D\tau$ at $N=101$.
	    a) $T=6\cdot10^{-4}$~K, the inequality~(\ref{eq:20}) yields the region of pair entanglement (k+1=2), the region is above the horizontal line; 
	    b) $T=3.2\cdot10^{-4}$~K, the region of the many-spin entanglement is a strip bounded by the horizontal lines with~$k=19$~and~$k=46$; 
	    c) $T = 1.6\cdot10^{_4}$~K, the horizontal lines ($k=18$ and k=$86$) bound the strip with many-spin entanglement;
	    d) $T=4.8\cdot10^{-5}$~K, entangled clusters with $11-92$ spins emerge.
	}
	\label{fig:2}
\end{figure}

The considered model of the spin-carrying molecules (atoms) in a nanopore in the dipolar ordered states expands possibilities of the investigation of many-spin entanglement in comparison with the related model~\cite{Doronin_2019},
in which the system was initially in the thermodynamic equilibrium in a strong external magnetic field.
The model~\cite{Doronin_2019}  is not useful for the investigation of the time-evolution of the system
because the stationary distribution of MQ NMR coherences establishes very quickly~\cite{Doronin_2009}.
Many-spin entanglement changes with the temperature in a very narrow temperature interval in the model~\cite{Doronin_2019}.
For example, all spins are entangled in the system consisting of 201 spins at the temperature $T=6.856\cdot10^{-3}$~K~\cite{Doronin_2019}.

The time dependence of the quantum Fisher information in the system consisting of 101 spins is presented in Fig.~(\ref{fig:2}) at different temperatures. 
One can see from Fig.~(\ref{fig:2}a) that only pair entanglement exists at the temperature $T=6\cdot10^{-4}$~K.
At the temperature $T=3.2\cdot10^{-4}$, one can see a strip in Fig.~(\ref{fig:2}b), in which the inequality~(\ref{eq:20}) can be satisfied when $19 \leq k \leq 46$.
Thus, there is many-spin entanglement in spin clusters consisting of 20-47 spins at the temperature $3.2\cdot10^{-4}$~K.
When the temperature decreases, the width of the strip, in which many-spin entanglement exists, increases. 
At the temperature $T=1.6\cdot10^{-4}$~K (Fig.~(\ref{fig:2}c)), clusters of 19-87 entangled spins emerge, and at the temperature $T=4.8\cdot10^{-5}$~K (Fig.~(\ref{fig:2}d)), we have 11-92 entangled spins.

\begin{figure}
  	\includegraphics[width=0.95\linewidth]{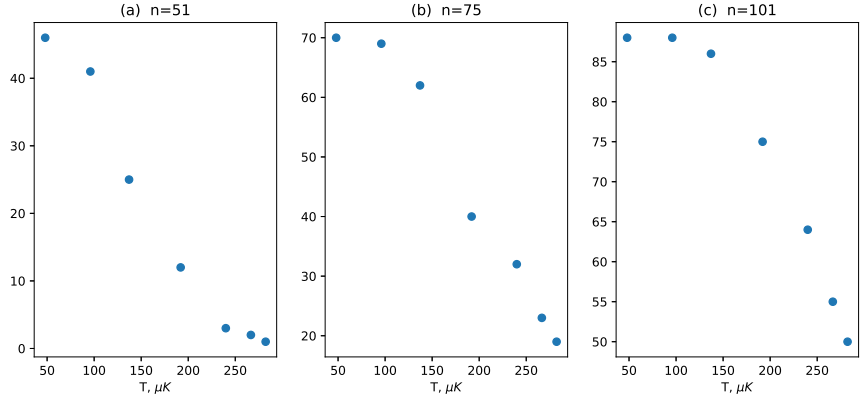}
	\caption{
	    The dependence of the maximal number of the entangled spins,
	    averaged over the evolution time $(0 \leq D\tau \leq 3)$, 
	    on the temperature at a) $N=51$; b) $N=75$; c) $N=101$.
	}
	\label{fig:3}
\end{figure}

The dependence of the average of the maximal number of the entangled spins over the evolution time $({0}\leq \mathrm{D}\tau\leq{3})$ on the temperature at different numbers of the spins in a nanopore is presented in Fig.~(\ref{fig:3}).
The maximal number of the entangled spins decreases when the temperature increases. 
The maximal number of the entangled spins increases when the number of the spins in the nanopore increases, because the system in the nanopore gets denser.

\section{Conclusion}
\label{sec:6}

We investigated many-spin entanglement in a system of spin-carrying molecules (atoms) filling a non-spherical nanopore in the conditions of the MQ NMR spectroscopy. The spins are in the dipolar ordered state initially.
We found the dependence of many-spin entanglement on the temperature and the number of the spins in the nanopore.

We believe that the MQ NMR spectroscopy is a subtle and useful method for the investigation of different quantum information problems.
In particularly, it is a very effective method for the exploration of quantum entanglement.

\section{Acknowledgements}
This work was performed as a part of a state task, State Registration No. 0089-2019-0002. 
This work was partially supported by the Russian Foundation for Basic Research (Grants Nos. 20-03-00147, 19-32-80004). 
I.L. acknowledges support from the Advancement of Theoretical Physics and Mathematics BASIS No. 19-1-5-130-1.

\appendix
\section{The two-pulse Broekaert-Jeener experiment at a low Zeeman temperature and a high dipolar temperature.}

Initially the system is in the thermodynamic equilibrium state in the strong external magnetic field with the density matrix
\begin{equation}
    \label{eq:a1}
   \sigma_{i} = \dfrac{e^{\beta_\mathrm{L} \omega_{0} I_\mathrm{z}}}{Z_{i}} ,
   \quad
   Z_{i} = \mathrm{Tr}\left\{e^{\beta_\mathrm{L} \omega_{0} I_\mathrm{z}} \right\}
\end{equation}
After the first resonance rf x-pulse, one has
\begin{equation}
    \label{eq:a2}
    \sigma'(0) = e^{ i \frac \pi 2 I_\mathrm{x}}
    \sigma_{i}
    e^{-i \frac \pi 2 I_\mathrm{x}}
    = \dfrac{e^{\beta_\mathrm{L} \omega_{0} I_\mathrm{y}}}{Z_{i}}  .
\end{equation}
Then the system evolves freely during the time $\tau$, 
and one applies the second resonance y-pulse rotating spins by angle $\theta$ around the y-axis of the RRF.
As a result, one obtains that
\begin{equation}
    \label{eq:a3}
    \sigma'(\tau) 
    = \dfrac{
      e^{-i \theta I_\mathrm{y}} e^{-i H_\mathrm{dz} \tau} 
      e^{\beta_\mathrm{L} \omega_{0} I_\mathrm{y}}
      e^{i H_\mathrm{dz} \tau} e^{i \theta I_\mathrm{y}}
    }{Z_{i}}. 
\end{equation}
After the time $T_2$ ($T_2$ is the spin relaxation time \cite{Goldman_1970}) the system achieves the thermodynamic equilibrium state
\begin{equation}
    \label{eq:a4}
    \sigma_{f} 
    = \dfrac{ e^{\alpha \omega_{0} I_\mathrm{z} + \beta H_\mathrm{dz}} }{Z_f},
\end{equation}
where $\alpha$ and $\beta$ are the inverse Zeeman and dipolar temperatures.
It is evident that the system has a single equilibrium state 
and there is a unique choice of the temperatures $\alpha$ and $\beta$ 
that is consistent with the conservation laws.
Those temperatures can be obtained from the conservation laws
\begin{align}
    \label{eq:a5}
    \mathrm{Tr} \left\{ I_\mathrm{z} \sigma'(\tau) \right\}
    & = \mathrm{Tr} \left\{ I_\mathrm{z} \sigma_{f}(\tau) \right\}
    \\
    \label{eq:a6}
    \mathrm{Tr} \left\{ H_\mathrm{dz} \sigma'(\tau) \right\}
    & = \mathrm{Tr} \left\{ H_\mathrm{dz} \sigma_{f}(\tau) \right\}
\end{align}
One can rewrite $\mathrm{Tr} \left\{ I_\mathrm{z} \sigma'(\tau) \right\}$ as 
\begin{multline}
    \label{eq:a7} 
    \tr{I_\mathrm{z} \sigma'(\tau)}
    =  \dfrac{1}{Z_{i}} \tr{
        e^{i \theta \sy} \sz e^{-i \theta \sy}
        e^{-i \hdz \tau} e^{\beta_\mathrm{L} \omega_{0} \sy} e^{i \hdz \tau}
    } 
    \\
    = \dfrac{1}{Z_i} \tr{
        \left( \cos(\theta) \sz - \sin(\theta) \sx \right) 
        e^{-i \hdz \tau} e^{\beta_\mathrm{L} \omega_{0} \sy} e^{i \hdz \tau}
    }
    \\
    = \dfrac{1}{Z_i} \tr{
        e^{-i \pi \sy}  
        \left( \cos(\theta) \sz - \sin(\theta) \sx \right) 
        e^{-i \hdz \tau} e^{\beta_\mathrm{L} \omega_{0} \sy} e^{i \hdz \tau}
        e^{i \pi \sy}  
    }
    \\
    = - \dfrac{1}{Z_i} \tr{
        \left( \cos(\theta) \sz - \sin(\theta) \sx \right) 
        e^{-i \hdz \tau} e^{\beta_\mathrm{L} \omega_{0} \sy} e^{i \hdz \tau} 
    } = 0
\end{multline}
In~(\ref{eq:a7}) we took into account that $\left[ e^{-i \pi \sy}, \hdz \right] = 0$.
Since we consider the case of a high dipolar temperature, it is possible for rewrite~(\ref{eq:a5}) as
\begin{equation}
    \label{eq:a8}
    0 = \dfrac{1}{Z_f} \tr{ \sz e^{\alpha \omega_0 \sz}}
    + \dfrac{\beta}{Z_f} \tr{\sz e^{\alpha \omega_0 \sz} \hdz}.
\end{equation}
Notice that $\tr{\sz} = \tr{\sz\hdz} = 0$. It means that $\alpha = 0$ satisfies Eq.~(\ref{eq:5}). 
Thus, we obtain the dipolar ordered state in the considered case.

\bibliography{bibliography}
\end{document}